\documentclass[prb,superscriptaddress,twocolumn,floatfix,showpacs]{revtex4-1}
\usepackage{amsmath,graphicx,latexsym,natbib,bm,psfrag,color}

\begin{document}

\title{Nonlinear optical properties of TeO$_2$ crystalline phases 
       from first principles}

\author{ Nabil Berka\"ine}
\affiliation{ Sciences des procds ceramiques et traitements de surface, 
              University of Limoges, Limoges 87000, France}

\author{Emmanuelle Orhan}
\affiliation{ Sciences des procds ceramiques et traitements de surface, 
              University of Limoges, Limoges 87000, France}

\author{Olivier Masson}
\affiliation{ Sciences des procds ceramiques et traitements de surface, 
              University of Limoges, Limoges 87000, France}

\author{Philippe Thomas}
\affiliation{ Sciences des procds ceramiques et traitements de surface, 
              University of Limoges, Limoges 87000, France}

\author{ Javier Junquera }
\affiliation{ Departamento de Ciencias de la Tierra y
              F\'{\i}sica de la Materia Condensada, Universidad de Cantabria,
              Avda. de los Castros s/n, 39005 Santander, Spain}

\date{\today}

\begin{abstract}
 We have computed second and third nonlinear optical susceptibilities
 of two crystalline bulk tellurium oxide polymorphs: 
 $\alpha$-TeO$_{2}$ (the most stable crystalline bulk phase) and
 $\gamma$-TeO$_{2}$ (the crystalline phase that ressembles the more to the 
 glass phase).
 Third order nonlinear susceptibilities of the crystalline phases
 are two orders of magnitude larger than $\alpha$-SiO$_{2}$ cristoballite,
 thus extending the experimental observations on glasses to the case
 of crystalline compounds.
 While the electronic lone pairs of Te contribute to those large values, 
 a full explanation of the anisotropy of the third order susceptibility
 tensor requires a detailed analysis of the structure, in particular 
 the presence of helical chains, that seems to be linked to cooperative
 non-local polarizabilty effects.
 Our results demonstrate that first-principles simulations are a 
 powerful predictive tool to estimate nonlinear optical susceptibilitites
 of materials.
\end{abstract}


%
\maketitle

\section{Introduction}
\label{section:introduction}

 Tellurium oxide glasses arouse lots of interest in the field 
 of nonlinear optics (NLO) since
 their unusual nonlinear optical indices have been noticed.
 The third order optical susceptibilities $\left[ \chi^{(3)} \right]$ 
 exhibited by pure TeO$_{2}$ glasses,
 of the order of 14 $\times$ 10$^{-13}$ esu, \cite{Kim-93} are
 indeed among the highest observed for oxide glasses 
 (50 times larger than in pure silica glasses) and they thus
 are of great interest in both fundamental science and technological 
 applications as optical modulators and frequency 
 converters.\cite{Hall-89,Kim-93,Mallawany} 

 The origin of these high values is not fully established yet. 
 Using a combination of experimental techniques (interferometric
 measurements) and {\it ab initio} calculations
 (within the restricted Hartree-Fock scheme), some authors~\cite{Fargin-96}  
 suggested that the highly polarizable  
 $5s^{2}$ electronic lone pair of Te$^{{\rm (IV)}}$
 could be responsible for the high NLO indices.
 Other theoretical works,\cite{shigeru1,shigeru2}
 based on the use of hybrid functionals within the 
 density functional theory (DFT), 
 reinforced this idea by demonstrating the importance of the 
 Te$^{{\rm (IV)}}$ lone pair on the hyperpolarizabilities of  
 {\it isolated} TeO$_{4}$ and TeO$_{3}$ structural units.
 This conclusion was extrapolated, by extension, to the case
 of TeO$_{2}$ based glasses.
 However, another series of theoretical 
 studies,\cite{Noguera-03,Noguera-04,mathsou06,Soulis-08}
 also carried out in the framework of DFT with
 hybrid functionals on (XO$_{2}$)$_{n}$ (X = Si or Te)
 polymer clusters of different shapes (chains, rings and cage geometries)
 and sizes (monitored by the number $n$ of XO$_{2}$ units)
 suggested another origin for the unusually 
 high values of the NLO susceptibilities, highlighting 
 the relevance of the structural features themselves,
 in particular how the structural blocks (i. e. TeO$_{2}$ units)
 are linked together. Only one type of such molecules, the linear chains, 
 seem to be capable of realistically reproducing the high 
 hypersusceptibility values for the tellurium oxides. This was
 attributed to an exceptionally strong nonlocality of the electronic
 polarization in these chains, much more important for the Te than for
 the Si oxides.

 Nevertheless, previous works were based on hypothetical fragments
 that were supposed to be likely found in the glass.
 The question about the high nonlinear susceptibility in the solid
 phases was not addressed.
 Clearly, further studies are needed to achieve a deeper
 understanding in the origin of these properties and notably to gauge
 the relative importance of the lone pair versus the structural features.
 A new way to tackle this problem is to treat the case of TeO$_{2}$-based
 crystalline compounds. 
 This would prevent the recourse to hypothetical structural fragments
 to feed the first-principles calculations. Besides,
 it would allow to study how the anisotropic nature in the crystalline
 phases translates into the variations of dielectric susceptibilities
 with crystalline directions.

 Unfortunately, the situation for crystals is different than for molecules 
 and two main problems arise in the first-principles 
 calculation of the hypersusceptibilities.
 The optical susceptibilities are derivatives of the 
 bulk macroscopic polarization with respect to electric field.
 If the polarization can easily be expressed in terms of the charge 
 distribution for molecules (finite systems), 
 it can not be obtained that way for crystals 
 (infinite systems treated periodically). The polarization in a 
 periodic system would indeed depend on the choice of 
 the unit cell.\cite{Martin-74}
 Solutions arised in the early 1990s and are often referred to as the 
 ``modern theory of polarization".\cite{King-Smith-93} The basic idea 
 is to consider the change in polarization \cite{Resta-92} of a crystal as 
 it undergoes some slow change, e.g. a slow displacement 
 of one sublattice relative to the others, and relate it to 
 the {\it current} that flows during this adiabatic 
 evolution of the system.\cite{Resta-94}
 The second problem lies in the nature of the applied electric 
 field that is macroscopic. 
 The scalar potential of a macroscopic homogeneous electric field is
 non-periodic (so the Bloch theorem does not apply), 
 and unbounded from below (so the energy of the system can always be 
 lowered transferring electrons to regions sufficiently far away, 
 hampering the aplicability of traditional variational methods).\cite{Souza-02}
 The first approach to circumvent this problem in first-principles simulations,
 due to Kunc and Resta,\cite{Kunc-83}
 was to consider ``sawtooth'' potentials in a supercell.
 We have previously tested this scheme, as implemented in the {\sc Crystal06} 
 program, \cite{crystal-web} on the computation of hypersusceptibilities
 of TeO$_{2}$ crystalline oxides.\cite{BenYahia-08} Although this method gives 
 satisfactory results, it requires defining a supercell for 
 keeping the periodicity along the applied field direction. 
 The dimension of the studied system is thus 
 very quickly limiting in terms of expansiveness in time and 
 computational requirements.
 A more recent variational alternative, 
 firmly rooted on the modern theory of polarization,
 was due to Souza, \'I\~niguez and Vanderbilt.\cite{Souza-02,Souza-04} 
 It is based on the 
 minimization of a electric enthalpy functional with respect
 to a set of polarized Bloch functions, thus including the effect of
 the electric field directly inside the unit cell.
 This approach, recently implemented in the {\sc Siesta} 
 method \cite{Soler-02,siesta-web} is the one used in the present work.
 
 In this paper we compute the second and third order optical susceptibility
 tensors in two {\it bulk} TeO$_{2}$ polymorphs:
 the $\alpha$-TeO$_2$ phase (known as paratellurite) that is the most 
 stable one,
 and the $\gamma$-TeO$_2$ phase,
 whose structure ressembles the more to the glass.
 The estimations of the nonlinear susceptibility data provided in the present
 work intend to fill the gap in the reported values of these quantities.
 Unfortunately, there is a cruel lack of experimental nonlinear susceptibility
 data for crystalline phases, mainly due to the difficulty of growing
 sufficiently large single crystals.
 To our knowledge, only the $\chi^{(2)}$ susceptibility tensor
 elements for the $\alpha$-TeO$_{2}$ phase has been measured, while
 there are no experimental values of the $\chi^{(3)}$ tensor elements for
 any crystalline phase.
 In addition, third order susceptibility tensor of 
 $\alpha$-SiO$_2$-cristobalite, which is structurally similar 
 to $\alpha$-TeO$_2$, is computed for comparison.

 The rest of the paper is organized as follows. 
 After presenting the computational details in 
 Sec. \ref{section:computationaldetails},
 and the structure characteristics of the different polymorphs 
 in Sec. \ref{section:structure},
 we describe the methodology used to compute the nonlinear 
 susceptibilities in Sec. \ref{section:nlop}.
 Second-order susceptibility values are then calculated
 in Sec. \ref{sec:second-order},
 and compared to experimental results in order to test the
 validity and the limitations of the method.
 Finally, the method is used as a predictive tool through 
 the calculation of the third order susceptibilities
 in Sec. \ref{sec:third-order},
 and clues are given for exploring relevant features 
 responsible for large variations of the dielectric susceptibilities
 with crystalline directions.

\section{Computational details}
\label{section:computationaldetails}

 We have carried out density functional first-principles simulations based
 on a numerical atomic orbital method as implemented in the {\sc Siesta} 
 code. \cite{Soler-02}
 All the calculations have been carried out within the 
 Generalized Gradient Approximation (GGA), using the functional
 parametrized by Perdew, Burke and Ernzerhof (PBE) \cite{Perdew-96}
 to simulate the electronic exchange and correlation. 

 Core electrons were replaced by {\it ab initio} norm conserving 
 pseudopotentials, generated using the 
 Troullier-Martins scheme, \cite{Troullier-91} in the 
 Kleinman-Bylander fully non-local separable representation. \cite{Kleinman-82}
 The $5s$ and $5p$ electrons of Te,
 $2s$ and $2p$ electrons of O, and
 $3s$ and $3p$ electrons of Si were considered as valence electrons
 and explicitly included in the simulations.
 In order to avoid the spiky oscillations close to the nucleus
 that oftenly appear in GGA-generated pseudopotentials,
 we have included small partial core corrections \cite{Louie-82} 
 for all the atoms.
 Te pseudopotential was generated scalar relativistically.
 The reference configuration, cutoff radii for each angular momentum shell, 
 and the matching radius between the full core charge density and the
 partial core charge density for the
 non-linear-core-corrections (NLCC) for the pseudopotentials used
 in the present work 
 can be found in Table \ref{table:pseudopotentials}. 

 \begin{table*}
    \caption[ ]{ Reference configuration, cutoff radii, and matching
                 radius between the full core charge and the
                 partial core charge for the
                 pseudopotentials used in our study. Units in bohr.
               }
    \begin{center}
       \begin{tabular}{ccccc}
          \hline
          \hline
                        &
                        &
          Te            &
          Si            &
           O            \\
          Reference                         &
                                            &
          $5s^{2}, 5p^{4}, 5d^{0}, 4f^{0}$  &
          $3s^{2}, 3p^{2}, 3d^{0}, 4f^{0}$  &
          $2s^{2}, 2p^{4}, 3d^{0}, 4f^{0}$  \\
          \hline
          Core radius                       &
          $s$                               &
          2.00                              &
          1.77                              &
          1.15                              \\
                                            &
          $p$                               &
          2.00                              &
          1.96                              &
          1.15                              \\
                                            &
          $d$                               &
          3.00                              &
          2.11                              &
          1.15                              \\
                                            &
          $f$                               &
          3.00                              &
          2.11                              &
          1.15                              \\
          Matching radius NLCC              &
                                            &
          1.30                              &
          1.50                              &
          1.17                              \\
          Scalar relativistic?              &
                                            &
          yes                               &
           no                               &
           no                               \\
          \hline
          \hline
       \end{tabular}
    \end{center}
    \label{table:pseudopotentials}
 \end{table*}

 The one-electron Kohn Sham eigenvalues were expanded in a basis
 of strictly localized \cite{Sankey-89} 
 numerical atomic orbitals. \cite{Artacho-99, Soler-02}
 The size of the basis set chosen was double-$\zeta$ plus polarization 
 for the valence states
 of all the atoms. All the parameters that define the shape 
 and the range of the basis functions were obtained by a variational
 optimization of the energy in the $\alpha$-cristobalite 
 polymorph of SiO$_{2}$, and of the enthalpy (with a pressure P = 0.2 GPa) 
 in the $\alpha$-phase of TeO$_{2}$, following the recipes given in 
 Refs. \onlinecite{Junquera-01} and \onlinecite{Anglada-02}.
 In both cases the optimization of the basis set 
 was performed at the experimental
 lattice parameters and internal positions taken from 
 Ref. \onlinecite{Thomas-88} for $\alpha$-TeO$_{2}$ and
 from Ref. \onlinecite{Yong-Il-05} for $\alpha$-cristobalite.
 
 The electronic density, Hartree, and exchange correlation potentials,
 as well as the corresponding matrix elements between the basis orbitals,
 were calculated in a uniform real space grid. An equivalent plane
 wave cutoff of 400 Ry was used to represent the charge density.
 During the geometry optimizations, 
 we used a $6 \times 6 \times 6$ Monkhorst-Pack mesh \cite{Monkhorst-76} 
 for all the Brillouin zone integrations. 
 The macroscopic polarization and its derivatives
 with respect to an external electric field, depend highly on 
 the number of $k$-points used. 
 To quantify this dependence, we have refined the Monkhorst-Pack meshes and 
 followed the evolution of the field induced 
 polarization with increasing number of $k$-points.
 Further details will be given in Sec. \ref{section:nlop-convergence}.
 
 For the structural characterization in the absence of an external electric
 field, the atoms were allowed to relax until the maximum component of the
 force on any atom was smaller than 0.01 eV/\AA, and the 
 maximum component of the stress tensor was smaller than 0.0001 eV/\AA$^{3}$.

\section{Structural properties of the crystalline phases}
\label{section:structure}

 \subsubsection{$\alpha$-SiO$_2$ cristobalite}
 \label{section:structurecrist}	

\begin{figure}[ht!]
    \begin{center}
       \includegraphics[width=\columnwidth] {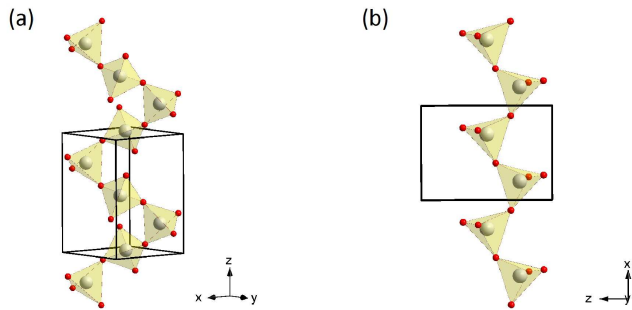}
       \caption{(Color online) Schematic view of the 
                $\alpha$-SiO$_2$ cristobalite unit cell 
                in different perspectives.
                Si and O  atoms are represented by grey and red balls,
                respectively. Solid lines mark the unit cell.
                (a) highlights the helical chains formed
                by the tetrahedra along $z$, while
                in (b) the zig-zag chains in the $x$ direction are 
                clearly observed. }
       \label{fig:SiO2-fig}
    \end{center}
 \end{figure}

 $\alpha$-SiO$_2$ cristobalite 
 crystallizes in the same space group ($P4_12_12$, $D^{4}_{4}$, no. 92)
 and with the same independent atomic positions as paratellurite,
 $\alpha$-TeO$_{2}$.
 Two atoms are independent by symmetry: one Si atom at position
 ($u$,$u$,0), and one O atom at position ($x$,$y$,$z$).
 The unit cell contains four formula units (twelve atoms).
 In the case of $\alpha$-cristobalite, the SiO$_4$
 entities are almost regular tetrahedra (see Fig. \ref{fig:SiO2-fig})
 and the Si-O distances are 
 all  close to 1.60 \AA\ (see Table-\ref{table:structuralsio2}).
 As shown in Fig.~\ref{fig:SiO2-fig},
 the tetrahedra are organized as to form an helical chain 
 along the $z$-direction. While in the directions $x$ and $y$, 
 the structure is made by zig-zag chains.

 \begin{table*}
    \caption[ ]{ Lattice constants (in \AA), and Wyckoff structural 
                 parameters for $\alpha$ cristobalite SiO$_{2}$ 
                 (space group $P4_{1}2_{1}2$).
                 PW stands for a plane wave method with pseudopotentials. 
                 d(Si-O)$_{1}$ and d(Si-O)$_{2}$ represent
                 the Si-O bond lengths inside the tetrahedra.
                 PBE stands for the 
                 Perdew-Burke and Ernzerhof \cite{Perdew-96} generalized
                 gradient functional, and 
                 B3LYP stands for a three-parameter hybrid functional 
                 (including part of the exact HF exchange.~\cite{B3LYP}).
                 For the O-Si-O angle the average between the four 
                 possible values is shown.
               }
    \begin{center}
       \begin{tabular}{cccccccc}
          \hline
          \hline
                                                                          &
          {\sc Siesta}                                                    &
          PW              \footnote{Reference \onlinecite{Coh-08}.}       &
          All electron    \footnote{Reference \onlinecite{Catti-00}.}     &
          All electron    \footnote{Reference \onlinecite{BenYahia-08}.}  &
          Expt.           \footnote{Reference \onlinecite{Yong-Il-05}.}   &
          Expt.           \footnote{Reference \onlinecite{Pluth-85}.}     \\
          xc-functional                                                   &
          PBE                                                             &
          PBE                                                             &
          B3LYP                                                           &
          B3LYP                                                           &
                                                                          &
                                                                          \\
          \hline
          \multicolumn{7}{c}{Cell parameters (\AA) }                      \\
          $a$                & 
          4.994              &
          5.073              &
          4.989              &
          5.028              &
          4.983              &
          4.957              \\
          $c$                &
          6.936              &
          7.085              &
          6.902              &
          7.013              &
          6.955              &
          6.890              \\
          \multicolumn{7}{c}{Atomic positions}                          \\
          Si($u$)            &
          0.305              &
          0.300              &
          0.307              &
          0.299              &
          0.306              &
          0.305              \\
          O($x$)             &
          0.236              &
          0.238              &
          0.236              &
          0.240              &
          0.251              &
          0.238              \\
          O($y$)             &
          0.118              &
          0.108              &
          0.119              &
          0.104              &
          0.095              &
          0.111              \\
          O($z$)             &
          0.186              &
          0.182              &
          0.186              &
          0.178              &
          0.156              &
          0.183              \\
          \multicolumn{7}{c}{Bond lengths (\AA)}                       \\
          d(Si-O)$_{1}$      &
          1.629              &
          1.646              &
          1.629              &
          1.615              &
          1.535              &
          1.600              \\
          d(Si-O)$_{2}$      &
          1.638              &
          1.646              &
          1.632              &
          1.626              &
          1.606              &
          1.620              \\
          \multicolumn{7}{c}{Angles (deg)}                             \\
          Si-O-Si            &
          142.00             &
          144.39             &
          142.28             &
          146.58             &
          158.86             &
          144.55             \\
          $<$ O-Si-O $>$     &
          109.32             &
          109.73             &
          109.78             &
          109.75             &
          119.96             &
          109.39             \\ 
          \hline
          \hline
       \end{tabular}
    \end{center}
    \label{table:structuralsio2}
 \end{table*}

 Theoretical lattice parameters and Wickoff positions are reported in
 Table-\ref{table:structuralsio2}, together with some experimental
 values for comparison. 
 Although the data summarized in Table-\ref{table:structuralsio2}
 include results obtained with different implementations of the 
 density functional theory (differences in the electrons included
 explicitly in the calculation, with and without pseudopotentials,
 different basis sets, different ways of sampling the Brillouin zone),
 a general trend that can be observed is that both
 PBE-GGA and B3LYP hybrid functional yield an 
 overestimation of the lattice constant of up to 3 \% with respect to
 the experimental values. 
 The Wyckoff positions obtained with the different 
 DFT-methods are in very good agreement between them
 [the maximum difference being in the O($y$) position],
 and are perfectly comparable with the experimental results.
 Indeed, this good performance in a traditionally complicated
 system like SiO$_{2}$ (very sensible to many approximations,
 in particular to basis sets \cite{Junquera-01}) 
 validates the {\sc Siesta} basis sets
 and pseudopotentials used in the present work. 
 
 \subsubsection{$\alpha$-TeO$_{2}$ phase}  
 \label{section:structurealpha}	

\begin{figure}[ht!]
    \begin{center}
       \includegraphics[width=\columnwidth] {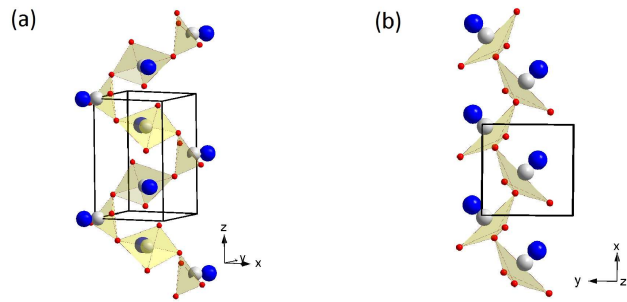}
       \caption{(Color online) Schematic view of the 
                $\alpha$-TeO$_2$ unit cell in different perspectives.
                Te and O  atoms are represented by grey and red balls,
                respectively, while the blue spheres are the lone pairs  
                of the Te atoms. 
                Solid lines mark the unit cell.
                Meaning of the panels as in Fig.~\ref{fig:SiO2-fig}.}
       \label{fig:aTeO2}
    \end{center}
 \end{figure}

 $\alpha$-TeO$_2$ crystallizes in a tetragonal unit cell 
 with the space group symmetry $P4_12_12$ ($D^{4}_{4}$, no. 92)
 as discussed before. \cite{Thomas-88}
 An schematic view of the unit cell is depicted in Fig. \ref {fig:aTeO2}.

 The tellurium atom occupies the center of triangle bipyramids 
 whose basis is formed by two oxygen atoms 
 and by the tellurium electronic lone pair, and whose apexes are also 
 oxygen atoms. Therefore, the Te atoms is coordinated with 
 four O atoms. This TeO$_4$ bypiramidal unit, building block of the
 tellurium oxides discussed in the present work, is referred to as
 a disphenod. Two different Te-O bond can be distinguished within
 the disphenod, being the equatorial O atoms closer to Te than the apical
 O atoms (experimental distances of 1.87 \AA\ and 2.12 \AA , respectively).
 As in $\alpha$-SiO$_2$
 cristobalite, the polyhedra are connected by vertices to form a three
 dimensional network, ressembling to an helical chain along $z$
 direction and zig-zag chains along $x$ and $y$ directions
 (see Fig. \ref{fig:aTeO2}.)

 Unit cell lattice parameters and internal coordinates 
 are reported in Table-\ref{table:structure-aTeO2}.
 The structural parameters obtained with {\sc Siesta} 
 are in very good agreement with those obtained with a plane wave code
 with ultrasoft pseudopotentials and an energy cutoff of 30 Ry,
 showing the good performance of the basis set used in the present work.
 As usual, the standard overstimation of the experimental equilibrium volume 
 by the generalized gradient approximation is found.
 The calculated independent bond-lengths at the theoretical
 equilibrium lattice parameters also overstimates the experimental 
 numbers. Nevertheless, the difference between the short equatorial 
 and the long axial Te-O bond lengths is preserved within {\sc Siesta}.

 \begin{table}
    \caption[ ]{ Lattice constants and Wyckoff structural 
                 parameters for paratellurite $\alpha$-TeO$_{2}$ 
                 (space group $P4_{1}2_{1}2$, $D^{4}_{4}$, no. 92).
                 PW stands for a plane wave calculation performed 
                 with the Quantum-Espresso package. \cite{espresso-web}
                 O$_{eq}$ and O$_{ap}$ represent,
                 respectively, the equatorial and apical 
                 oxygens within the disphenods.
                 Both {\sc Siesta} and plane wave simulations have been
                 carried out within the PBE-GGA functional.\cite{Perdew-96}
                 All electron results were computed with the B3LYP hybrid
                 functional.\cite{B3LYP}
                 Units of the lattice constants and distances in \AA\ and 
                 angles in degrees.
               }
    \begin{center}
       \begin{tabular}{ccccc}
          \hline
          \hline
                                                                          &
          {\sc Siesta}                                                    &
          PW              \footnote{Reference \onlinecite{Ceriotti-06}.}  &
          All electron    \footnote{Reference \onlinecite{BenYahia-08}.}  &
          Expt.           \footnote{Reference \onlinecite{Thomas-88}.}    \\
          \hline
          \multicolumn{5}{c}{Cell parameters (\AA) }                      \\
          $a$                & 
          4.987              &
          4.990              &
          4.899              &
          4.808              \\
          $c$                &
          7.606              &
          7.546              &
          7.792              &
          7.612              \\
          \multicolumn{5}{c}{Atomic positions}                          \\
          Te($u$)            &
          0.0261             &
          0.0272             &
          0.0276             &
          0.0268             \\
          O($x$)             &
          0.1418             &
          0.1467             &
          0.1389             &
          0.1368             \\
          O($y$)             &
          0.2494             &
          0.2482             &
          0.2585             &
          0.2576             \\
          O($z$)             &
          0.1973             &
          0.1968             &
          0.1845             &
          0.1862             \\
          \multicolumn{5}{c}{Bond lengths (\AA)}                       \\
          d(Te-O$_{\rm eq}$) &
          1.955              &
          1.944              &
          1.909              &
          1.879              \\
          d(Te-O$_{\rm ap}$) &
          2.192              &
          2.118              &
          2.160              &
          2.121              \\
          \multicolumn{5}{c}{Angles (deg)}                             \\
          O$_{\rm eq}$-Te-O$_{\rm eq}$  &
          104.6                         &
          103.6                         &
          103.2                         &
          103.4                         \\
          O$_{\rm ap}$-Te-O$_{\rm ap}$  &
          169.8                         &
          171.2                         &
          168.1                         &
          168.0                         \\
          Te-O-Te                       &
          136.2                         &
          137.1                         &
          139.1                         &
          138.6                         \\         
          \hline
          \hline
       \end{tabular}
    \end{center}
    \label{table:structure-aTeO2}
 \end{table}

 \subsubsection{$\gamma$-TeO$_{2}$ phase}  
 \label{section:structuregamma}	

\begin{figure}[ht!]
    \begin{center}
       \includegraphics[width=\columnwidth] {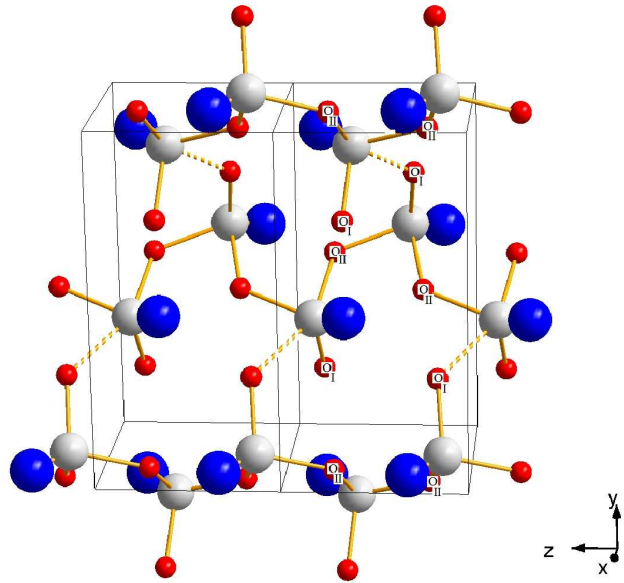}
       \caption{(Color online) 
                Schematic view of the $\gamma$-TeO$_2$ unit cell.
                Te and O  atoms are represented by grey and red balls,
                respectively, while the blue spheres are the lone pairs  
                of the Te atoms. 
                Solid lines mark the unit cell, repeated along the 
                $z$-direction for the sake of clarity. 
                The yellow dashed lines are the longest
                Te-O distance.
                }
       \label{fig:gTeO2}
    \end{center}
 \end{figure}

\begin{figure}[ht!]
    \begin{center}
       \includegraphics[width=\columnwidth] {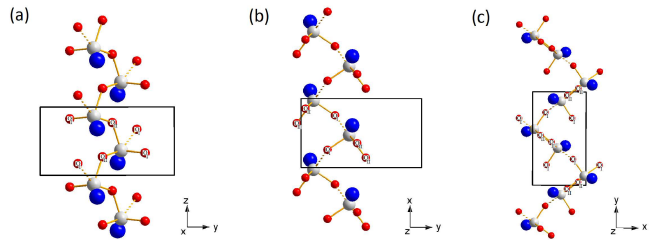}
       \caption{(Color online) 
                Schematic view of the $\gamma$-TeO$_2$ unit cell in different
                perspectives.
                Te and O  atoms are represented by grey and red balls,
                respectively, while the blue spheres are the lone pairs  
                of the Te atoms. 
                Solid lines mark the unit cell.
                The two symmetry  inequivalent O atoms are labeled 
                as O$_{I}$ and O$_{II}$.
                The yellow dashed lines are the longest
                Te-O distance.
                (a) and (b) highlight the zig-zag chain along the
                $z$ and $x$ direction, respectively.
                (c) focus on the helical chain along the $y$ direction.
                }
       \label{fig:gTeO2-chains-fig}
    \end{center}
 \end{figure}

 $\gamma$-TeO$_2$ crystallizes in 
 an orthorhombic unit cell with the space group 
 $P2_12_12_1$ ($D^{4}_{2}$, no. 19).\cite{Champarnaud-00} 
 A schematic view of the unit cell is represented in 
 Fig. \ref{fig:gTeO2}. 
 This phase is metastable at normal conditions, 
 and has been recently identified by x-ray powder diffraction
 of recrystallized amorphous TeO$_{2}$ doped with oxides. 
 The unit cell contains four formula units (twelve atoms), with 
 three atoms independent by symmetry: one Te atom located at
 ($u$,$v$,$w$), and two oxygen atoms labeled as O$_{I}$ and O$_{II}$.
 As in the $\alpha$-phase, the structure of the $\gamma$-phase 
 can be considered as a 3D network of corner-sharing TeO$_4$ disphenods.
 However, in the $\gamma$-phase
 the disphenods are strongly deformed,
 so the length of the four Te-O bonds are rather different
 (experimentally the bond lengths range from 1.86 \AA\ to 2.20 \AA),
 with a much larger spread (0.34 \AA) than in the $\alpha$-phase (0.24 \AA).
 Indeed, if we assume that the longest Te-O distance (marked with a 
 dashed line in Fig. \ref{fig:gTeO2})
 is too long to form a chemical covalent bond, 
 then the structure can be viewed as 
 an infinite zig-zag chain of TeO$_3$ units in the $z$
 direction, connected by the bridge Te-O$_{II}$-Te. Including now the
 longest bond Te-O, the disphenoids TeO$_4$ forms zig-zag chains along $x$
 direction and helical chains along $y$ direction but with different
 bridges (Te-O$_{I}$-Te and Te-O$_{II}$-Te) 
 (see Fig. \ref{fig:gTeO2-chains-fig}.)

 In Fig.~\ref{fig:gTeO2}, despite the fact that the
 bridge between Te atoms along the $z$-oriented chains is always 
 through a O$_{II}$ atom, the
 length of the Te-O$_{II}$ bond is different, ranging between 1.94 \AA\
 for one of the bonds of the chain to 2.02 \AA\ for the second bond.

 \begin{table}
    \caption[ ]{ Lattice constants and Wyckoff structural 
                 parameters for $\gamma$-TeO$_{2}$ 
                 (space group $P2_{1}2_{1}2_{1}$, $D^{4}_{2}$, no. 19).
                 PW stands for a plane wave calculation performed 
                 with the Quantum-Espresso package. \cite{espresso-web}
                 d(Te-O)$_{1}$ and d(Te-O)$_{2}$ 
                 stand for the short and the long equatorial Te-O distance,
                 while d(Te-O)$_{1^{'}}$ and d(Te-O)$_{2^{'}}$
                 represent the long and the short axial bond lengths
                 within the disphenod. 
                 Both {\sc Siesta} and plane wave simulations have been
                 carried out within the PBE-GGA functional.\cite{Perdew-96}
                 Units of the lattice constants and distances in \AA .
               }
    \begin{center}
       \begin{tabular}{cccc}
          \hline
          \hline
                                                                          &
          {\sc Siesta}                                                    &
          PW              \footnote{Reference \onlinecite{Ceriotti-06}.}  &
          Expt.           \footnote{Reference \onlinecite{Champarnaud-00}.}\\
          \hline
          \multicolumn{4}{c}{Cell parameters (\AA) }                      \\ 
          $a$                & 
          5.181              &
          5.176              &
          4.898              \\
          $b$                & 
          8.636              &
          8.797              &
          8.576              \\
          $c$                &
          4.446              &
          4.467              &
          4.351              \\
          \multicolumn{4}{c}{Atomic positions}                          \\
          Te($u$)            &
          0.9632             &
          0.9581             &
          0.9686             \\
          Te($v$)            &
          0.1034             &
          0.1032             &
          0.1016             \\
          Te($w$)            &
          0.1368             &
          0.1184             &
          0.1358             \\
          O$_{I}$($x$)       &
          0.7770             &
          0.7641             &
          0.759              \\
          O$_{I}$($y$)       &
          0.2935             &
          0.2851             &
          0.281              \\
          O$_{I}$($z$)       &
          0.1750             &
          0.1645             &
          0.173              \\
          O$_{II}$($x$)      &
          0.8616             &
          0.8599             &
          0.855              \\
          O$_{II}$($y$)      &
          0.0380             &
          0.0406             &
          0.036              \\
          O$_{II}$($z$)      &
          0.7259             &
          0.7131             &
          0.727              \\
          \multicolumn{4}{c}{Bond lengths (\AA)}                       \\
          d(Te-O)$_{1}$      &
          1.912              &
          1.900              &
          1.859              \\
          d(Te-O)$_{2}$      &
          1.983              &
          1.960              &
          1.949              \\
          d(Te-O)$_{1^{'}}$  &
          2.116              &
          2.119              &
          2.019              \\
          d(Te-O)$_{2^{'}}$  &
          2.315              &
          2.252              &
          2.198              \\
          \hline
          \hline
       \end{tabular}
    \end{center}
    \label{table:structure-gTeO2}
 \end{table}

 The theoretical lattice parameters and independent positions
 of the atoms are reported in Table-\ref{table:structure-gTeO2}.
 The good agreement between {\sc Siesta} and plane waves results 
 confirms and highlights the transferability of our basis set,
 that was optimized for the $\alpha$-TeO$_{2}$
 structure.
 Again, the PBE-GGA functional overstimates the experimental equilibrium
 volume, although the deviation in this case (9 \%) is slightly larger
 than usual.  This overstimation translates also in a slight
 overestimation of the spread of the bond lengths in 
 the case of {\sc Siesta} (0.40 \AA).

 Regarding the first-principles simulations on the structure
 of the TeO$_{2}$ phases, we can summarize that the 
 disphenoidal configuration of the TeO$_4$ entities are respected 
 both in the $\alpha$ and $\gamma$ phases.
 The lone pair sterical effect is thus conserved in our geometry 
 optimization. 
 The good comparison between our structural parameters and the ones obtained
 with plane waves \cite{Ceriotti-06} support the use of the numerical 
 atomic orbital method implemented in {\sc Siesta}
 in the present study.

 \section{Nonlinear optical properties}
 \label{section:nlop}

 \subsection{Methodology}
 \label{section:nlop-method}

 When an intense light (a powerful laser) goes through an insulator, 
 this medium responds nonlinearly. 
 Its polarization $\bm{\mathcal{P}}$ 
 can be expressed as a Taylor expansion 
 of the electric field $\bm{\mathcal{E}}$.

 \begin{align}
    \mathcal{P}_{i} = & \mathcal{P}_{i}^{s} + 
                        \sum_{j=1}^{3} \frac{\partial \mathcal{P}_{i}}
                                          {\partial \mathcal{E}_{j}} 
                                          \mathcal{E}_{j} + 
                        \sum_{j,k=1}^{3} \frac{1}{2}  
                                     \frac{\partial^{2} \mathcal{P}_{i}}
                                          {\partial \mathcal{E}_{j} 
                                           \partial \mathcal{E}_{k}}
                                          \mathcal{E}_{j} \mathcal{E}_{k} 
    \nonumber \\
                    & + \sum_{j,k,l=1}^{3} \frac{1}{6} 
                                     \frac{\partial^{3} \mathcal{P}_{i}}
                                          {\partial \mathcal{E}_{j} 
                                           \partial \mathcal{E}_{k}
                                           \partial \mathcal{E}_{l}}
                                          \mathcal{E}_{j} \mathcal{E}_{k}  
                                          \mathcal{E}_{l} +
                     \cdots,
   \label{eq:polarisation1}
 \end{align}

 \noindent where $i, j, k$ and $l$ refer to cartesian directions, and
 $\mathcal{P}_{i}^{s}$ is the zero-field (spontaneous) polarization.
 The coefficients of the previous expansion, derivatives of the polarization
 with respect to the electric field of increasing order, are related with 
 the electrical susceptibilities of the material. In particular, 
 the coefficient of the linear term is directly proportional to the 
 linear dielectric susceptibility (a second-order rank tensor),

 \begin{equation}
    \chi_{ij}^{(1)} = \frac{1}{\varepsilon_{0}} \frac{\partial \mathcal{P}_{i}}
                                                     {\partial \mathcal{E}_{j}},
    \label{eq:linearsusceptibility}
 \end{equation}
 
 \noindent where $\varepsilon_{0}$ is the permittivity of free space. 
 Unless otherwise is specified, we will use the SI system of units
 throughout the paper.
 In the same way, the coefficients of the quadratic and cubic terms are used to 
 define the second-order and third-order 
 nonlinear susceptibilities,

 \begin{equation}
    \chi_{ijk}^{(2)} = \frac{1}{2\varepsilon_{0}} 
                       \frac{\partial^{2}\mathcal{P}_{i}}
                            {\partial \mathcal{E}_{j} \partial \mathcal{E}_{k}},
    \label{eq:quadraticsusceptibility}
 \end{equation}

 \begin{equation}
    \chi_{ijkl}^{(3)} = \frac{1}{6\varepsilon_{0}} 
       \frac{\partial^{3}\mathcal{P}_{i}}
            {\partial \mathcal{E}_{j} \partial \mathcal{E}_{k} \mathcal{E}_{l}}.
    \label{eq:cubicsusceptibility}
 \end{equation}

 \noindent Replacing 
 Eqs. (\ref{eq:linearsusceptibility})-(\ref{eq:cubicsusceptibility})
 into Eq. (\ref{eq:polarisation1}), then the expansion of the polarization 
 can be written as

 \begin{align}
    \mathcal{P}_{i} = & \mathcal{P}_{i}^{s} + 
                      \sum_{j=1}^{3} \varepsilon_{0}
                                     \chi^{(1)}_{ij}
                                     \mathcal{E}_{j} +
                      \sum_{j,k=1}^{3} \varepsilon_{0}
                                     \chi^{(2)}_{ijk}
                                     \mathcal{E}_{j} \mathcal{E}_{k} 
    \nonumber \\
                 & +  \sum_{j,k,l=1}^{3} \varepsilon_{0}
                                     \chi^{(3)}_{ijkl}
                                     \mathcal{E}_{j} \mathcal{E}_{k}  
                                     \mathcal{E}_{l} +
                     \cdots
   \label{eq:polarisation2}
 \end{align}

 In general, the polarization depends both on the valence electrons and 
 the ions. In the present work, we deal only with the electronic contribution.
 The main reason behind this approximation is that the second (SHG) and 
 third (THG) harmonic generation experiments 
 leading to the second and third order susceptibilities 
 are performed at frequencies high-enough to get rid of ionic relaxations
 but low enough to avoid electronic excitations. \cite{Veithen-05}
 This constraint implies that both
 the frequency of $\bm{\mathcal{E}}$ 
 and its second and third harmonics are
 lower than the fundamental adsorption gap.
 Indeed, SHG and THG experiments are pump-probe settings  
 which are sensitive only to the 
 electronic contributions. 

 For a completely anisotropic crystal, Eq. (\ref{eq:polarisation2}) 
 implies respectively 9, 27 and 81 elements for 
 $\chi^{(1)}$, $\chi^{(2)}$ and $\chi^{(3)}$ tensors. 
 These numbers can be reduced significantly by considering the 
 specific crystal class.\cite{Boyd} The nonvanishing elements
 for the symmetry groups of 
 $\alpha$-TeO$_{2}$ and $\gamma$-TeO$_{2}$ 
 can be found in Table-\ref{table:symmetryalphaTiO2}
 and Table-\ref{table:symmetrygammaTiO2}, respectively.
 
 \begin{table}[h!]
    \caption[ ]{ Symmetry-allowed components of the 
                 linear dielectric susceptibility, $\chi^{(1)}$,
                 and the second-order, $\chi^{(2)}$,
                 and third-order, $\chi^{(3)}$,
                 nonlinear optical susceptibilities tensors for
                 the $P4_{1}2_{1}2$ space group, in which both the
                 $\alpha$-TeO$_{2}$ and the $\alpha$-SiO$_2$ cristoballite
                 crystallize.
               }
    \begin{center}
       \begin{tabular}{c|c|c}
          \hline  
          \hline  
          $\chi^{(1)}$  & 
          $\chi^{(2)}$  & 
          $\chi^{(3)}$  \\ 
          \hline  
          $xx$                                       & 
          $xyz=-yxz$                                 &    
          $xxxx=yyyy$                                \\ 
          $yy$                                       &
          $xzy=-yzx$                                 &
          $zzzz$                                     \\
          $zz$                                       & 
          $zxy=-zyx$                                 &  
          $yyzz=xxzz$                                \\
                                                     & 
                                                     &  
          $yzzy=xzzx$                                \\
                                                     & 

                                                     &  
          $xxyy=yyxx$                                \\
                                                     & 
                                                     &   
          $yzyz=xzxz$                                \\
                                                     & 
                                                     &   
          $xyxy=yxyx$                                \\
                                                     & 
                                                     &   
          $zzyy=zzxx$                                \\
                                                     & 
                                                     &   
          $zyyz=zxxz$                                \\ 
                                                     & 
                                                     &   
          $zyzy=zxzx$                                \\ 
                                                     & 
                                                     &   
          $xyyx=yxxy$                                \\ 
          \hline 
          \hline  
       \end{tabular}
    \end{center}
    \label{table:symmetryalphaTiO2}
 \end{table}

 \begin{table}[h!]
    \caption[ ]{ Same as in Table-\ref{table:symmetryalphaTiO2}
                 but for the $P2_{1}2_{1}2_{1}$ space group, in which the
                 $\gamma$-TeO$_{2}$ phase crystallizes.
               }
    \begin{center}
       \begin{tabular}{c|c|c}
          \hline  
          \hline  
          $\chi^{(1)}$  & 
          $\chi^{(2)}$  & 
          $\chi^{(3)}$  \\ 
          \hline  
          $xx$                                                   & 
          $xyz$,$yxz$                                            & 
          $xxxx$, $yyyy$, $zzzz$, $yyzz$, $xxzz$, $yzzy$, $xzzx$ \\
          $yy$                                                   &
          $xzy$, $yzx$                                           & 
          $xxyy$, $yyxx$, $yzyz$, $xzxz$, $xyxy$, $yxyx$, $zzyy$ \\ 
          $zz$                                                   &   
          $zxy$, $zyx$                                           & 
          $zzxx$, $zyyz$, $zxxz$, $zyzy$, $zxzx$, $xyyx$, $yxxy$ \\
          \hline 
          \hline  
       \end{tabular}
    \end{center}
    \label{table:symmetrygammaTiO2}
 \end{table}

 Thus, combining the symmetry-allowed components of the susceptibility
 tensors with a judicious choice of the direction of the applied
 electric field, we can restrict the expansion in Eq. (\ref{eq:polarisation2})
 so that only a given component of the susceptibility is present.
 Then, its value can be obtained by fitting the polarization 
 versus electric field dependence as obtained from first-principles
 computations of the response of the bulk materials to an homogeneous
 electric field.\cite{Souza-02,Souza-04}

 To better fix this procedure, let us develop the way the 
 $\chi^{(2)}_{yxz}$ of $\gamma$-TeO$_2$ is computed.
 Taking into account the nonvanishing optical susceptibilities of its 
 crystal class, $P 2_1 2_1 2_1$, 
 the expansion of the Eq. (\ref{eq:polarisation2}) in this case 
 is reduced to

 \begin{align}
    \mathcal{P}_{y} = & \mathcal {P}^{s}_{y} + 
                        \varepsilon_0 \chi^{(1)}_{yy} \mathcal{E}_{y}  
    \nonumber \\
        & + \varepsilon_{0} \left(
            \chi^{(2)}_{yxz} \mathcal{E}_{x} \mathcal{E}_{z} + 
            \chi^{(2)}_{yzx} \mathcal{E}_{z} \mathcal{E}_{x} \right)    
    \nonumber \\
        & + \varepsilon_{0} \left( 
            \chi^{(3)}_{yyyy} \mathcal{E}_{y} \mathcal{E}_{y} \mathcal{E}_{y} +
            \chi^{(3)}_{yyzz} \mathcal{E}_{y} \mathcal{E}_{z} \mathcal{E}_{z} + 
            \chi^{(3)}_{yzzy} \mathcal{E}_{z} \mathcal{E}_{z} \mathcal{E}_{y} 
            \right)
    \nonumber \\
        & + \varepsilon_{0} \left(
            \chi^{(3)}_{yyxx} \mathcal{E}_{y} \mathcal{E}_{x} \mathcal{E}_{x} + 
            \chi^{(3)}_{yzyz} \mathcal{E}_{z} \mathcal{E}_{y} \mathcal{E}_{z} + 
            \chi^{(3)}_{yxyx} \mathcal{E}_{x} \mathcal{E}_{y} \mathcal{E}_{x} 
            \right)
    \nonumber \\
        & + \varepsilon_{0} \left( 
            \chi^{(3)}_{yxxy} \mathcal{E}_{x} \mathcal{E}_{x} \mathcal{E}_{y}
            \right).
    \label{eq:expanchiyxz}
 \end{align}

 \noindent To isolate the $\chi^{(2)}_{yxz}$ component, we can apply
 a field with only $x$ and $z$ components,
 $ {\bm{\mathcal{E}}} = 
 ( \mathcal{E}, 0, \mathcal{E} )$, so the 
 expansion in Eq. (\ref{eq:expanchiyxz}) reduces to 

 \begin{equation}
    \mathcal{P}_{y} = \mathcal{P}^{s}_{y} + 
                      \varepsilon_{0}  
                      \left( \chi^{(2)}_{yxz} + \chi^{(2)}_{yzx} \right)
                      \mathcal{E}^2.
    \label{eq:expanchiyxz2}
 \end{equation}

 \noindent Assuming the Kleinman symmetry conditions,\cite{kleinman}
 that states that the nonlinear optical susceptibility tensor, as 
 defined in Eq. (\ref{eq:quadraticsusceptibility}), is symmetric under
 a permutation of the $i, j, k$ indices so 
 $\chi^{(2)}_{yxz} = \chi^{(2)}_{yzx}$,
 then Eq. (\ref{eq:expanchiyxz2}) transforms into 

 \begin{equation}
    \mathcal{P}_{y} = \mathcal{P}^{s}_{y} + 
                      2 \varepsilon_{0}  \chi^{(2)}_{yxz} \mathcal{E}^2.
    \label{eq:expanchiyxz3}
 \end{equation}

 Up to now we have applied only basic definitions and symmetry properties.
 The ingredient from first-principles comes from the computation
 of the field-induced electronic polarization 
 (the ionic cores are clamped at the equilibrium zero-field positions)
 as a function of the external electric field.
 Here, we have used
 the method of Refs. \onlinecite{Souza-02} and \onlinecite{Souza-04}. 
 In these milestone works, 
 Souza, \'I\~niguez and Vanderbilt showed how to compute
 stationary states of a periodic system at a finite, constant electric
 field $\bm{\mathcal{E}}$ 
 on a uniform discrete $k$-point mesh,
 providing that the magnitude of the field does not exceed a critical value
 $\mathcal{E}_{c}(N)$, that decreases as the number of $k$-points
 $N$ increases. 
 The algorithm, described in Sec. V of Ref. \onlinecite{Souza-04},
 is based on the diagonalization of a field-dependent Hermitian operator,
 $\hat{T}_{\bf{k}} (\bm{\mathcal{E}})$ in the notation of 
 Ref. \onlinecite{Souza-04}, for every $k$-point in the mesh.
 The field-dependent operator depends explicitly on a given $k$,
 but implicitly couples neighboring $k$-points.
 Therefore, to know the occupied Bloch-like eigenstates, the diagonalization
 has to be iterated until the procedure converges at all the 
 $k$-points and the occupied subspace stabilizes.
 Once self-consistency has been achieved, the overlap matrix between
 the cell periodic parts of the Bloch functions at neighboring $k$-points
 can be used to compute the component of the polarization parallel to a vector
 of the reciprocal lattice, as usual in the discretized formulation
 of the polarization of a solid as a Berry-phase. \cite{King-Smith-93}

\begin{figure}[ht!]
    \begin{center}
       \includegraphics[width=\columnwidth] {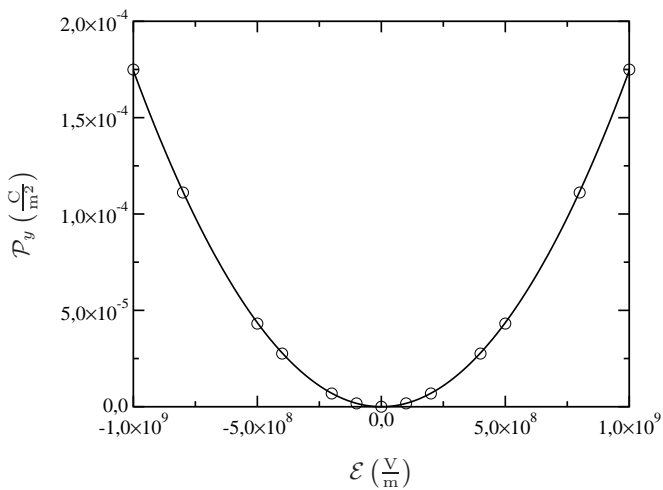}
       \caption{ Field-induced polarization along $y$, $\mathcal{P}_{y}$,
                 as a function of an applied electric field only
                 with components along 
                 the $x$ and $z$ directions, 
                 $\bm{\mathcal{E}}= 
                 (\mathcal{E}, 0, \mathcal{E})$,
                 for the $\gamma$-TeO$_2$ phase. 
                 The $x$-axis represents the component of the field
                 along the $x$-direction. 
                 Atomic coordinates and unit cell lattice parameters
                 are clamped to the optimized structure at zero external field,
                 and only the electronic response is considered while
                 computing the polarization.
                 Circles are the first-principles results 
                 and the solid line is the fit to Eq. (\ref{eq:expanchiyxz3}).}
       \label{fig:fit} 
    \end{center}
 \end{figure}

 Once a set of polarization versus electric field data have been obtained
 from first principles, one can make a choice between different methods
 to extract a value for the nonlinear susceptibility.
 The first one is a direct fit of the raw data 
 (see the symbols of Fig. \ref{fig:fit})
 to expressions like Eq. (\ref{eq:expanchiyxz3}).
 
 \psfrag{Figure1x}[cc][cc]{$\mathcal{E} \left( \frac{\rm{V}}{\rm{m}} \right)$}
 \psfrag{Figure1y}[cc][cc]{$\mathcal{P}_{y} 
                            \left( \frac{\rm{C}}{\rm{m^{2}}} \right)$}

 \noindent The second one is to compute the derivatives of the
 macroscopic polarization with respect the electric field by finite differences,
 using the Richardson's extrapolation to estimate the limit $h \rightarrow 0$
 from calculations with two different step sizes:

 \begin{equation}
    f^{(n)} (x) = \frac{4}{3} D^{(n)} (x,h) - \frac{1}{3} D^{(n)} (x,2h)
                  + O(h^{4}), 
    \label{eq:richardson1}
 \end{equation}

 \noindent where $D^{(1)}$ is given by the centered finite difference 
 expresion 

 \begin{equation}
    D^{(1)} (x,h) \equiv \frac{f(x+h) - f(x-h)}{2h} = f^{'} (x) + O (h^{2}),
    \label{eq:richardson2}
 \end{equation}

 \noindent $D^{(2)}$ is given by

 \begin{equation}
    D^{(2)} (x,h) \equiv \frac{f(x+h) + f(x-h) - 2 f(x)}{h^{2}} = 
            f^{''} (x) + O (h^{2}),
    \label{eq:richardson3}
 \end{equation}

 \noindent and $D^{(3)}$ is given by

 \begin{align}
    D^{(3)} (x,h) \equiv & \frac{f(x+2h) -2 f(x+h) + 2 f(x-h) - f(x-2h)}
            {2h^{3}} 
    \nonumber \\ 
            = & f^{'''} (x) + O (h^{2}).
    \label{eq:richardson4}
 \end{align}

 \noindent Here, in the results quoted below, we have used
 a field step of $h$ = 0.04 V/\AA. 

 Comparison between the results obtained with the two methods
 will be presented in the next Sec. \ref{section:nlop-convergence}.

 \subsection{Convergence studies}
 \label{section:nlop-convergence}

 It is well known that the total energy ground state calculations
 {\it for insulators} converge expeonentially fast with respect
 to $k$-point sampling.
 However, while a given $k$-point sample might be perfectly 
 adequate for some properties, it might constitute too severe
 approximation for others. \cite{Mattson-05}
 In other words, the convergence with respect the $k$-point sampling
 is property-dependent. 
 That is the case of the computation of the polarization and its derivatives,
 when a discretized Berry phase polarization expression is used.\cite{Roman-06}
 In the formalism developed in Ref. \onlinecite{Souza-04},
 and summarized in Sec. \ref{section:nlop-method},
 both the calculation of the  
 stationary states of the periodic system at a finite constant electric
 field, and the Berry-phase polarization is made 
 on a uniform discrete $k$-point mesh in the first-Brillouin zone,
 and the convergence for finite field simulations is considerably
 slower than in total energy or charge density
 calculations.\cite{Roman-06}
 The situation does not improve significantly when a perturbation
 approach is used instead of the finite field method.
 Previous first-principles simulations on the nonlinear optical
 susceptibilities in the framework of the density functional perturbation
 theory,\cite{Veithen-05} show that the convergence of the results
 is quite poor with the number of spetial $k$ points, at least
 when the discretization of the formula for the Berry phase of the 
 occupied bands is performed after the perturbation expansion
 (although the situation improves dramatically when the perturbation expansion
 is performed after the discretization).
 We can wonder how rapidly our results converge with respect to
 the $k$-point sampling. 
 
\begin{figure}[ht!]
    \begin{center}
       \includegraphics[width=\columnwidth] {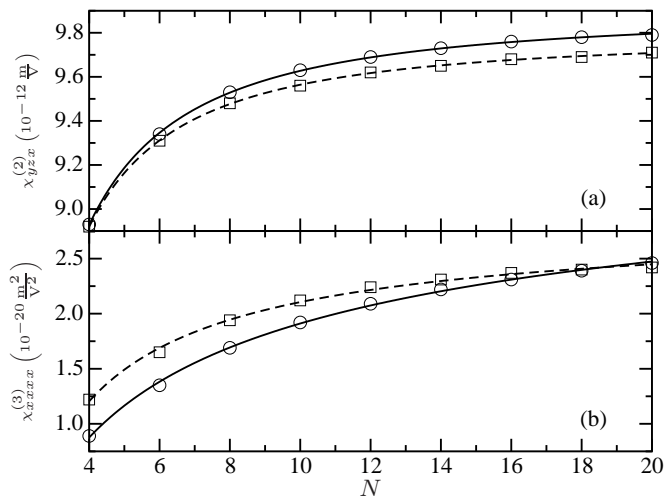}
       \caption{ (a) $\chi^{(2)}_{yxz}$ component of the second-order 
                 NLO susceptibility tensor of $\gamma$-TeO$_{2}$, 
                 and (b) the $\chi^{(3)}_{xxxx}$ component of the third-order 
                 NLO susceptibility tensor of $\alpha$-TeO$_2$ 
                 as a function of the size of the $N \times N \times N$ 
                 shifted Monkhorst-Pack \cite{Monkhorst-76} grid.
                 Circles and solid lines represent the results obtained with
                 a direct fit of the polarization versus electric field curve.
                 Squares and dashed lines have been obtained with the
                 Richardson extrapolation formula to compute derivatives
                 by finite differences. The lines are fits to analytical 
                 functions that would allow an extrapolation to 
                 $N \rightarrow \infty$.
                 }
       \label{fig:convergencek}
    \end{center}
 \end{figure}

 In Fig. \ref{fig:convergencek} we represent the 
 $\chi^{(2)}_{yxz}$ component of the second-order NLO susceptibility tensor
 of $\gamma$-TeO$_{2}$ [panel (a)], 
 and the $\chi^{(3)}_{xxxx}$ component of the third-order NLO susceptibility
 tensor of $\alpha$-TeO$_2$ [panel (b)] as a function of the size
 of the $N \times N \times N$ shifted Monkhorst-Pack \cite{Monkhorst-76} 
 grid. 
 The results are shown for the two different methods used to obtain 
 the derivatives of the macroscopic polarization with respect the
 electric field: the direct fit to the polarization versus field curve
 and the Richardson extrapolation to compute finite differences derivatives.
 The results provided by both methods are in good agreement, with 
 differences smaller than 2 \% for reasonable sizes of the
 Monkhorst-Pack mesh. For the rest of the paper all the reported results
 have been obtained with the Richardson extrapolation method.

 In any case, as we can see from Fig. \ref{fig:convergencek},
 the convergence with respect the number of $k$-points included
 in the simulations is rather slow. A way of predicting converged
 values\cite{Roman-06} for a given magnitude $p$ would be to explapolate to
 $N \rightarrow \infty$ through 
 a least square fit against an analytical formula 
 of the form $p = p_{\infty} + a/N^{b}$ (where
 $p_{\infty}$, $a$, and $b$ are adjustable parameters).
 However, this would require several calculations at different $N$.
 Here, we have proceed in a different way, computing only the
 values of the NLO susceptibilities for $N = 20$.
 This would lead to results with errors of usually around 
 2\% for the second-order and 
 up to 15 \% for the third-order susceptibilities.
 However, a similar trend is expected for estimations of the
 susceptibilities in different phases with approximately the 
 same unit cell volume and number of atoms per unit cell.
 Therefore, for the consequence of the main goal of this work, that is, 
 the {\it comparison} of the nonlinear optical susceptibilities between 
 phases to ascertain the origin of their large values,
 the previous approach is justified.
 
 \psfrag{Chi2figy}[cc][cc]{\tiny{$\chi^{(2)}_{yzx} 
                           \left( 10^{-12} \frac{\rm{m}}{\rm{V}} \right)$}}
 \psfrag{Chi3figy}[cc][cc]{\tiny{$\chi^{(3)}_{xxxx} 
                           \left( 10^{-20} \frac{\rm{m^2}}{\rm{V^{2}}}\right)$}}
 \psfrag{Labelxconvk}[cc][cc]{$N$}

 \section{Results}
 \label{sec:results}

 \subsection{Second order NLO susceptibilities}
 \label{sec:second-order}

 The values for the symmetry allowed components of the 
 second order susceptibilities of
 $\alpha$-TeO$_{2}$ and $\gamma$-Te$O_2$ are 
 gathered in Table-\ref{table:2NLO-results}, where they 
 are presented in terms of the $\boldsymbol{d}$ tensor,
 often used in the literature of nonlinear optics, and defined as 

 \begin{equation}
    d_{il}=d_{ijk}=\frac{1}{2} \chi^{(2)}_{ijk},
    \label{eq:defofd}
 \end{equation}
 
 \noindent where the first index, $i$, refers to a cartesian
 direction (1 for $x$, 2 for $y$, and 3 for $z$),
 while the second index $l$ contracts the two other cartesian
 indices $j$ and $k$ in Voigt notation (see Table \ref{table:Voigt}).

 \begin{table}[h!]
    \begin{center}
       \begin{tabular}{llcc}
          \hline
          \hline
          Material                       & 
          Element                        &    
          Theory                         & 
          Experiment                     \\
          \hline 
          $\alpha$-TeO$_2$               &  
          $d_{14}$                       & 
          $0.0$                          &  
          $1.4$ \footnote{Reference \onlinecite{Chemla-72}}-        
          $1.9$ \footnote{Reference \onlinecite{Singh-72}}     \\
                                         & 
          $d_{25}$                       & 
          $0.0$                          & 
                                         \\
              	                         &  
          $d_{36}$                       & 
          $0.0$                          &       
                                         \\
          $\gamma$-TeO$_2$               &  
          $d_{14}$                       & 
          $11.6$                         & 
                                         \\
                                         &
          $d_{25}$                       &
          $11.6$                         &  
                                         \\ 
                                         &  
          $d_{36}$                       & 
          $11.5$                         & 
                                         \\
          \hline
          \hline 
       \end{tabular}
       \caption{ Symmetry allowed values for the second order 
                 nonlinear susceptibilities of 
                 $\alpha$-TeO$_{2}$ and $\gamma$-TeO$_{2}$.
                 Units in 10$^{-9}$ esu
                 (1 esu = 4.192 $\times 10^{-4}$ m/V).}
       \label{table:2NLO-results}
    \end{center}
 \end{table}

 \begin{table}[h!]
    \begin{center}
       \begin{tabular}{cc}
          \hline
          \hline
          $l$        &
          $jk$       \\
          \hline
          1          &
          $xx$       \\
          2          &
          $yy$       \\
          3          &
          $zz$       \\
          4          &
          $yz = zy$  \\
          5          &
          $zx = xz$  \\
          6          &
          $xy = yx$  \\
          \hline
          \hline
       \end{tabular} 
       \caption{ Relationship between contracted index $l$
                 and cartesian indices $jk$ in the definition
                 of the $\boldsymbol{d}$ tensor in Eq. (\ref{eq:defofd}).
                 }
       \label{table:Voigt}
    \end{center}
 \end{table}

 According to (i) the conditions imposed by the space groups 
 (see Tables \ref{table:symmetryalphaTiO2} and \ref{table:symmetrygammaTiO2}),
 and (ii) the Kleinman symmetry, which means that
 $\chi^{(2)}_{ijk}$ is symmetric under a permutation
 of $i$, $j$, and $k$,
 all the elements of the NLO susceptibility tensor 
 of $\alpha$-SiO$_{2}$ and $\alpha$-TeO$_{2}$ should vanish,
 and all the independent elements of $\gamma$-TeO$_{2}$ should
 be equal.
 To illustrate this, let us take, for instance,
 the $d_{14}$ element for $\alpha$-TeO$_{2}$. The crystal symmetry imposes that

 \begin{equation}
    d_{14} = \frac{1}{2} \chi^{(2)}_{xyz} = 
           - \frac{1}{2} \chi^{(2)}_{yxz} = - d_{25},
    \label{eq:symcondd14}
 \end{equation}

 \noindent while following the Kleinman symmetry 
 
 \begin{equation}
    d_{14} = \frac{1}{2} \chi^{(2)}_{xyz} = 
             \frac{1}{2} \chi^{(2)}_{yxz} = d_{25}.
    \label{eq:kleincondd14}
 \end{equation}

 \noindent The only way to satisfy Eqs. (\ref{eq:symcondd14}) 
 and (\ref{eq:kleincondd14}) simultaneously is $d_{14} = d_{25} = 0$.
 A similar reasoning for $\gamma$-TeO$_{2}$, where in principle
 all the crystal symmetry allowed elements of the second order
 nonlinear susceptibility might be independent, shows
 that the Kleinman rule forces them to be equal.
 A permutation of indices impose that

 \begin{equation}
    d_{xyz} = d_{xzy} = d_{yxz} = d_{yzx} = d_{zxy} = d_{zyx}.
 \end{equation}

 \noindent Both results are clearly visible in Table \ref{table:2NLO-results}.
  
 Unfortunately, the comparison with the experiment is not straightforward.
 The only experimental results published up to now on crystals concerns 
 the $\alpha$-SiO$_{2}$ critoballite,\cite{Cabrillo-01} and
 $\alpha$-TeO$_2$ phase,\cite{porter} resulting in a weak SHG response
 (indicating the presence of a second order susceptibility)
 with the SHG efficiency of $\alpha$-TeO$_{2}$ five times larger 
 than in SiO$_{2}$. 
 The metastable nature of the $\gamma$-TeO$_2$ phase makes 
 impossible to grow single crystals big enough to 
 be optically characterized.  

 It can be surprising at first sight to obtain $\chi^{(2)}$ values for 
 the $\alpha$-TeO$_2$ where, as explained before, 
 the combination of the space group symmetry and the
 Kleinman's rules \cite{kleinman} should inhibit the presence of a SHG signal.
 Different mechanisms have been called to explain this
 apparent inconsistency. 
 First, Kleinman's relations are expected to breakdown
 if the second harmonic frequency is close to an absorption
 band of the material, but this was not the case for
 $\alpha$-TeO$_{2}$ under the measurement conditions.\cite{Chemla-72} 
 Second, although the Kleinman's rule is always satisfied 
 in non-dispersive media, it can be broken in dispersive materials, 
 as might be the case for the experimental $\alpha$-TeO$_{2}$ samples.
 As the present calculations were done in the static finite field
 limit, no dispersive effect can be taken into account so that the Kleinman's
 conditions are verified.

 \subsection{Third order NLO susceptibilities}
 \label{sec:third-order}

 A glance to Tables \ref{table:symmetryalphaTiO2} 
 and \ref{table:symmetrygammaTiO2} reveals that there are
 21 elements of the third order nonlinear susceptibility 
 allowed by spatial symmetry considerations.
 A first-principles estimation of all of the independent 
 elements would be a rather cumbersome task, that is out of
 the scope of this work.
 Instead, we have computed the 
 $xxxx$, $yyyy$, and $zzzz$ elements of the 
 third order susceptibilities,
 as they are informative enough for the 
 structure/properties relationship considerations.
 For those particular components both the polarization
 and the electric fields involved are directed along the
 same cartesian direction [see Eq. (\ref{eq:cubicsusceptibility})].
 Results for $\alpha$-SiO$_{2}$, 
 $\alpha$-TeO$_{2}$ and $\gamma$-TeO$_{2}$
 are summarized in Table-\ref{table:3NLO-results}.

 The results obtained with {\sc Siesta} using the variational
 implementation of the finite field in periodic system approach
 are in very good agreement with those obtained 
 with {\sc Crystal06}, where
 the field was introduced as a sawtooth potential. However, the last
 approach requires the use of a supercell to adapt the periodicity
 of the potential and, therefore, the increase of the computational cost
 of the simulation.

 Unfortunately, the comparison with the experiment is not
 so straightforward, since there are no measurements yet 
 concerning the crystalline phases of TeO$_{2}$ or SiO$_{2}$. 
 Therefore, the values reported in Table-\ref{table:3NLO-results} 
 can be considered as purely predictive.
 The only $\chi^{(3)}$ available data
 for tellurium oxides has been measured on the 
 corresponding glass. \cite{Kim-93}
 However, Raman spectroscopy measurements \cite{Noguera-04}
 have shown that the structure of $\gamma$-TeO$_{2}$ is close to the
 structure of this glass, so it is tempting
 to think that the order of magnitude of the third-order
 susceptibility should be the same in both the glass and
 the crystalline $\gamma$-TeO$_{2}$ phase.

 \begin{table}[h!]
    \begin{center}
       \begin{tabular}{ccccc}
          \hline
          \hline
          Material                                       &
          Element of $\chi^{(3)}$                        &
          \multicolumn{3}{c}{$\chi^{(3)}$}               \\
                                                         & 
                                                         & 
          This work                                      & 
          All electron \footnote{Reference \onlinecite{BenYahia-08}.}   &
          Expt.        \footnote{Reference \onlinecite{Kim-93}.}        \\
          \hline
          $\alpha$-SiO$_2$                               & 
          $xxxx$                                         &
          0.4                                            &
          0.52                                           &
          {\it 0.28}                                     \\ 
                                                         &
          $zzzz$                                         &
          0.7                                            &
          0.59                                           & 
                                                         \\
          $\alpha$-TeO$_2$                               & 
          $xxxx$                                         &
          17.3                                           &
          18.36                                          & 
                                                         \\
                                                         &
          $zzzz$                                         &
          31.3                                           &
          32.07                                          & 
                                                         \\
          $\gamma$-TeO$_2$                               & 
          $xxxx$                                         &
          12.2                                           &
                                                         & 
          {\it 14.1}                                     \\
                                                         &
          $yyyy$                                         &
          23.6                                           &
                                                         &   
                                                         \\
                                                         &       
          $zzzz$                                         &
          20.8                                           &
                                                         &
                                                         \\
          \hline 
          \hline 
       \end{tabular}
       \caption{ Symmetry allowed values for the third order 
                 nonlinear susceptibilities of $\alpha$-SiO$_{2}$ cristoballite,
                 $\alpha$-TeO$_{2}$ and $\gamma$-TeO$_{2}$.
                 The all electron simulations have been carried out with
                 the B3LYP functional\cite{B3LYP} as implemented in 
                 the {\sc Crystal06} package.\cite{crystal-web}
                 The experimental values, written in italics, 
                 correspond to the corresponding glass phase, whose structure
                 for the Te oxide is not so different from 
                 the $\gamma$-TeO$_{2}$ phase as suggested by Raman 
                 spectroscopy measurements.\cite{Noguera-04}
                 Units in 10$^{-13}$ esu
                 (1 esu = 1.398 $\times 10^{-8}$ m$^{2}$/V$^{2}$).}
       \label{table:3NLO-results}
    \end{center}
 \end{table}

 As can be drawn from Table-\ref{table:3NLO-results}, the third order
 susceptibilities of TeO$_{2}$ and SiO$_{2}$ 
 crystals obtained from our first-principles
 simulations are of the same order of magnitude as the experimental
 values for the relevant glasses. 
 The very large $\chi^{(3)}$ predicted for the two tellurium oxide 
 polymorphs, on the range of $10^{-12}$ esu, are nearly two order of
 magnitudes larger than for crystallized silica,
 with an average
 $\chi^{(3)}$ (TeO$_{2}$) / $\chi^{(3)}$ (SiO$_{2}$) ratio close to 50,
 thus extending the experimental observations on glasses to the case
 of crystalline compounds.

 We can wonder now about the origin of the large values for $\chi^{(3)}$ 
 in TeO$_{2}$. 
 Despite the fact that $\alpha$-SiO$_{2}$ and $\alpha$-TeO$_{2}$
 crystallizes in the same space group, their third order
 susceptibilities are different by two orders of magnitude. 
 This fact points out that structural arrangement by itself
 can not be responsible for the remarkable NLO properties of TeO$_{2}$.

 A more detailed analysis of the $\chi^{(3)}$ tensor elements
 reveals that the highest value are for the $z$ direction 
 in the case of $\alpha$-SiO$_{2}$ and $\alpha$-TeO$_{2}$,
 and for the $y$ direction for the $\gamma$-TeO$_{2}$.
 As stated in Sec. \ref{section:structure}, those are precisely
 the crystalline directions where the chains display an helical shape.
 Moreover, for the two structures that 
 are structurally similar
 ($\alpha$-SiO$_{2}$ and $\alpha$-TeO$_{2}$)
 nearly the same ratio $\chi^{(3)}_{zzzz}$/$\chi^{(3)}_{xxxx}$
 of around 1.8 is obtained.
 Since no lone pair is present in SiO$_2$ compounds,
 Te lone pair effect (orientation of the LP with 
 respect to the electric field for example) 
 cannot be responsible for these large variations.
 
 This suggests that high $\chi^{(3)}$ values are structurally induced 
 and that helical chains are much more favourable 
 than the zig-zag chains structures
 shown along the $x$ direction for these materials.
 Indeed, Mirgorodsky {\it et al.}\cite{mathsou06} 
 and Soulis and coworkers~\cite{Soulis-08} have shown how
 there is a strong link between the structure of polymer TeO$_{2}$ molecules 
 and its nonlinear optical properties, with the chain-like species
 developing a drastic increase in their 
 third-order hyperpolarizability with increasing chain length.
 The Te-O-Te bridges play a dominant role in the polarization properties
 of the long TeO$_{2}$ chains.
 This was attributed to an exceptionally strong nonlocality of the 
 electronic polarization, that is, assuming that the electric field
 applied at a given point would induce a dipole moment not only at the 
 very point but in the vicinity of this point (extending up to eight-ten
 neighbors from the point of perturbation). 
 
 Now, let us turn our atention to the zig-zag chains
 of the two TeO$_{2}$ compounds (along the $x$ and $y$ direction for
 the $\alpha$-TeO$_{2}$ phase and along the $x$ and $z$ direction
 for the $\gamma$-TeO$_{2}$ phase.)
 They are all very similar in shape and made of 
 asymmetric single Te-O-Te bridges with the bond length
 sequence $-S-L-S-L-$ (where $S$ and $L$ stand for short and long bond lengths,
 respectively). We can define an asymmetry index as
 $AI = 100 \times (L-S)/L$, whose value amounts to 11
 for the chains parallel to $x$ or $y$ in $\alpha$-TeO$_{2}$,
 15 parallel to $x$ in $\gamma$-TeO$_{2}$,
 and 4 parallel to $z$ in $\gamma$-TeO$_{2}$.
 It is interesting to note that the $AI$ values evolve
 as the inverse of the $\chi^{(3)}$ values, 
 suggesting that, for a given chain, the more symmetrical 
 the bridge the higher the third order nonlinear optical susceptibility value.
 
\section{Conclusion}

 The second and third order nonlinear optical susceptibilities of
 two bulk crystalline phases of TeO$_{2}$, and $\alpha$-SiO$_{2}$
 cristoballite have been computed using the variational approach
 to compute the response of a periodic solid to an external electric field.

 The third order nonlinear susceptibilities are in good agreement 
 with previous more expensive theoretical predictions, were 
 the electric field is introduced by means of a sawtooth potential,
 and with the experimental results for related glass phases.
 Indeed we were could reproduce the large values for the $\chi^{(3)}$
 tensor elements of the tellurium oxides, 50 times larger than in pure silica
 glasses.
 
 Our results provide some clues to explain the origin
 for the high hypersusceptibilities and the large
 variations with respect to the crystalline directions.
 In particular these properties could be attributed to 
 the presence of helical chains in the structure.
 
 A next step to be taken in order to through some light into 
 the origin of the large
 values of the nonlinear optical susceptibilities would require
 the calculation of the center of the localized Wannier functions, 
 and their variation with the external fields.

 Our results demonstrate how first-principles calculations are
 an efficient tool to estimate 
 nonlinear optical susceptibilities of crystalline solids.
 These might contribute to fill the gap usually 
 left by experimental measurements due to the difficulty in growing 
 single crystals big enough to be optically characterized.

\begin{acknowledgments}
 The authors are indebted to Cecilia Pola for useful discussions on 
 the use of Richardson extrapolation method, and with 
 Vincent Rodr\'{\i}guez for illuminating us in the 
 the Kleinman symmetries.
 JJ acknowledges finantial support of the Spanish Ministery of Science and
 Innovation through the MICINN Grant FIS2009-12721-C04-02.
 Calculations were performed on the computers
 of the ATC group at the Universidad de Cantabria 
 and on the ``Calculateur Limousin'' cluster of the Universit\'e de Limoges.
\end{acknowledgments}

\end{document}